\documentstyle[aasms4,epsfig]{article}
 
\newcommand\beq{\begin{equation}}
\newcommand\eeq{\end{equation}}

\received{}
\accepted{}
 
\begin{document}
{\title{Spectral Energy Distributions and Age Estimates of 172 
Globular Clusters in M31}
 
\author{
Linhua Jiang\altaffilmark{1,2},
Jun Ma\altaffilmark{1},
Xu Zhou\altaffilmark{1},
Jiansheng Chen\altaffilmark{1},
Hong Wu\altaffilmark{1},
and
Zhaoji Jiang\altaffilmark{1}
}
 
\altaffiltext{1}
{National Astronomical Observatories of CAS and
CAS-Peking University joint Beijing Astrophysical Center,
Beijing, 100012, P. R. China }
 
\altaffiltext{2}
{Department of Astronomy,
Peking University, Beijing, 100871, P. R. China }
 
\begin{abstract}                 

In this paper we present CCD multicolor photometry for 172 globular clusters 
(GCs), taken from the Bologna catalog (Battistini et al. 1987), in the nearby spiral 
galaxy M31. The observations were carried out by using the National Astronomical 
Observatories $60/90$ cm Schmidt Telescope in 13 intermediate-band filters, which 
covered a range of wavelength from 3800 to 10000{\AA}. This provides a multicolor 
map of M31 in pixels of 
$1\arcsec{\mbox{}\hspace{-0.15cm}.} 7 \times 1\arcsec{\mbox{}\hspace{-0.15cm}.} 7$. 
By aperture photometry, we obtain the spectral energy distributions (SEDs) for 
these GCs. Using the relationship 
between the BATC intermediate-band system used for the observations and the 
{\it UBVRI} broad-band system, the magnitudes in the {\it B} and {\it V} bands are 
derived. The computed {\it V} and $B-V$ are in agreement with the values 
given by Battistini et al. (1987) and Barmby et al. (2000). 
Finally, by comparing the photometry of each GC with theoretical stellar population 
synthesis models, we estimate ages of the sample GCs for different metallicities. 
The results show that nearly all our sample GCs have ages more than $10^{9}$ years, 
and most of them are around $10^{10}$ years old. Also, we find that GCs fitted by 
the metal-poor model are generally older than ones fitted by the metal-rich model.

\end{abstract}                   

 
\section{INTRODUCTION} 

The study of globular clusters (GCs) plays an important role in our 
understanding of the evolution and structure of galaxies. A single 
globular cluster is a ``packet'' of population II stars with a single 
age and chemical abundance, so the GCs provide a unique laboratory 
for exploring dynamics and theoretical evolution of their parent 
galaxies. GCs are likely to be the oldest known stellar objects, and 
all large galaxies appear to contain them (Harris 1991). Among the Local Group 
galaxies, M31 is an ideal target for studying GCs, since it comprises 
the largest sample of GCs, which is more than all GCs combined in other 
Local Group members (Battistini et al. 1987; Racine 1991; Harris 1991; Fusi 
Pecci et al. 1993). It is thus well known that the understanding of M31 GCs is 
especially important.

M31, a large spiral galaxy, is one of our nearest neighbors, with a distance 
modulus of 24.47 (Holland 1998; Stanek \& Garnavich 1998). Because it is so near, 
and has a very large sample of GCs, M31 is an excellent candidate for GCs studies. 
The first catalog of 140 clusters in M31 was given by Hubble (1932). 
From then on, numerous lists of GC candidates were published. A general 
catalog of almost all the GCs known in the early time was compiled by Vete\u{s}nik 
(1962a), who gave a total number of 257 GCs, including most candidates 
identified by previous works (e.g. Hubble 1932; Seyfert \& Nassau 1945; Hiltner 
1958; Mayall \& Eggen 1953; Kron \& Mayall 1960). Later, several major catalogs 
were compiled by Sargent et al. (1977), Crampton et al.(1985), and Bologna 
Group (Battistini et al. 1980, 1987). Sargent et al. (1977) found or 
confirmed 355 GCs in their catalog, and also rejected 52 candidates obtained 
by previous surveys. The catalog of Crampton et al. (1985) was a compilation 
with 509 GC candidates, and the core radii, {\it V} magnitudes, 
and intrinsic colors for most candidates were also given or 
estimated. The most comprehensive catalog of globular cluster candidates 
may be the Bologna catalog (Battistini et al. 1987). Bologna Group did independent 
searches for candidates and compiled them with their own Bologna number. 
Bologna catalog contains a total of 827 objects, and all the objects were 
classified into five classes by authors' degree of confidence. 353 of these 
candidates were considered as class A or class B by high level of 
confidence, and the others fell into class C, D, or E. {\it V} magnitude and  
$B-V$ color for most candidates were also given in the Bologna catalog.

All globular cluster candidates in these catalogs were sure carefully selected and 
usually identified by several steps. However, it is not to say that 
these samples were clean. They were often contaminated by foreground stars, 
background galaxies, or other objects (e.g. HII regions). For example, even 
several class A, or B candidates in Bologna catalog turned out to be not 
GCs (Barmby et al. 2000). 
Similarly, these catalogs were not complete, although they may be fairly complete 
down to $V=18$ ($M_v \sim -6.5$) (Fusi Pecci et al. 1993). 
Actually, most candidates in Bologna catalog also reached down to $V=18$ 
(Battistini et al. 1987). So in recent years, many works were searching for new 
and fainter GCs, especially in the center region of M31(e.g. Auri\`{e}re, 
Coupinot, \& Hecquet 1992; Battistini et al. 1993). Additionally, some much fainter 
GCs were detected by {\it HST} with high spatial resolution (e.g. Barmby \& Huchra 
2001a; Rich et al. 2001). According to the estimate of Barmby \& Huchra (2001a), 
the total number of GCs in M31 is about $460\pm70$. 

Anyway, these samples provide a good database for studies of M31 GCs. 
On the basis of the database, many results were derived and some properties of M31 
GCs were well explored, such as luminosity function (e.g. Auri\`{e}re, Coupinot, 
\& Hecquet 1992; Mochejska et al. 1998; Barmby, Huchra, \& Brodie 2001b), reddening 
and intrinsic colors (e.g. Vete\u{s}nik 1962b; Bajaja \& Gergely 1977; 
Iye \& Richter 1985; Barmby et al. 2000), metallicities (e.g. van den Bergh 
1969; Ashman \& Bird 1993; Barmby et al. 2000; Perrett et al. 2002), and comparisons 
with the Galactic GCs and M33 GCs (e.g. Hiltner 1960; Frogel, Persson, \& Cohen 
1980; Reed, Harris, \& Harris 1992; Mochejska et al. 1998).

Although there are already much observation information and theoretical 
results in previous studies, for full physical information about the GCs, it 
is of great value to obtain multiband photometry, which can provide their accurate 
SEDs. In addition, we can estimate GC ages by comparing
their intrinsic SEDs with ones of theoretical stellar population synthesis
models. The globular cluster ages are important, they can provide
us with information on the early formative stages of the parent galaxy
and can be utilized to provide a lower limit to the age of the parent galaxy.
Especially, the distribution of
GC ages can be used to understand the conditions for their
formation. For example, Barmby \& Huchra (2000) presented that conditions
for cluster formation must have existed for a substantial fraction of
the parent galaxy's lifetime by comparing three sets of population
synthesis models with integrated colors of M31 and Galactic globular
clusters in set of $UBVRIJHK$.
However, ages of most individual GCs in M31 remain undetermined, 
although they have been derived for a few ones. For example, Jablonka, Alloin, \& 
Bica (1992) presented spectrophotometric data for 7 GCs, and derived their 
reddening, metallicities and ages. According to the calculation of Borges and Idiart 
(1995), de Freitas Pacheco (1997) reported a sample of 12 GCs, and obtained the mean 
age of the sample.

In this paper we present CCD spectrophotometry of a set of GCs in M31 
using images obtained with the Beijing-Arizona-Taiwan-Connecticut (BATC) 
Multicolor Sky Survey Telescope designed to obtain SED information for 
galaxies (Fan et al. 1996). The BATC system uses the 60/90 cm Schmidt telescope 
with 15 intermediate bandwidth filters. In this study we use 13 of these 
filters, from 3800 to 10000{\AA}, with images covering the most visible extent 
of M31, and present the SEDs of 172 GCs selected from the Bologna catalog 
(Battistini et al. 1987). Using previously derived relationships between the BATC 
intermediate-band system, we go on to compute the {\it B} and {\it V} magnitudes for 
these objects. The computed {\it V} magnitudes and $B-V$ colors 
are in good agreement with the values given by Battistini et al. (1987) and Barmby 
et al. (2000). We also plot histograms of the {\it V} band magnitudes and of 
$B-V$ colors, as well as plotting the corresponding color-magnitude diagram for 
our sample GCs. Since reddening of GCs in M31 was determined by several authors 
(e.g. Vete\u{s}nik 1962b; Crampton et al. 1985; Barmby et al. 2000), we can 
infer the intrinsic colors for each individual GC. Finally, using theoretical 
stellar population synthesis models, we estimate ages of our sample GCs. The results 
show that most objects are as old as about $10^{10}$ years, and GCs fitted by 
the metal-poor model are generally older than ones fitted by the metal-rich model.

This paper will take the following form: Section 2 presents our sample GCs, and 
details of observations and data reduction. We also obtain SEDs, and give comparisons 
of GC photometry between the BATC system and previous measurements in this
section. In Section 3, we briefly describe the stellar population synthesis models 
of GSSP. We present how we deredden the sample GCs in Section 4.
In Section 5, ages for the GCs are estimated for different values of metallicity. 
Finally, we give a summary in Section 6.

\section{SAMPLE OF GLOBULAR CLUSTERS, OBSERVATIONS AND DATA REDUCTION}
 
\subsection{Selection of Sample} 

The GC candidates in each catalog were sure carefully selected
and identified by different ways or several steps, however, the catalogs of
GCs were often contaminated by stars, background galaxies, or other objects
such as HII regions. The basic idea of the identification of a new GC candidate
is the fact that the point spread function (PSF) of a GC has a larger FWHM than
a star (Mochejska et al. 1998). In general, people discover a new GC first using
visual inspection, i.e., distinguish cluster images from stars or spurious
non-stellar objects by image morphology (Battistini et al. 1987). For example,
objects that appear asymmetrical or exhibit structure are usually not GCs, thus
most open clusters and background galaxies which are asymmetrical or display
structure are easy to be eliminated (Crampton et al. 1985). Sometimes identifying
a new GC in the center region or in the halo of M31 need other means.
Due to the strong background gradient, the identification of candidates in the
center region by image morphology may be less efficient. Auri\`{e}re et al. (1992)
used another powerful method, the wavelet analysis, to reduce the image.
For the candidates in the M31 halo, there are also several other ways to
define a clean sample (see Reed, Harris, \& Harris 1992 for a detail).

The sample of GCs chosen in this paper is from the Bologna catalog by
Battistini et al. (1987). As has been mentioned, the Bologna catalog with
a total of 827 GC candidates, is the most comprehensive catalog of M31
GCs. All the candidates were classified into five classes by authors' degree
of confidence. 353 of all the candidates were considered as class A (254 objects)
and class B (99 objects) by very high and high level of confidence, 152
candidates were listed in class C, and the others fell into class D and E.
Our selection of the sample is as follows. First, we take all candidates of class
A and B (i.e. Table IV of Bologna catalog) as our original sample. We find that
only 223 objects are in our CCD field. Then we note that cluster 7, 55, 132 and
147 are virtually stars, not true GCs (Barmby et al. 2000), so they are not
included in our sample, i.e., there are 219 class A or B GCs of Bologna catalog in
our CCD field. Besides, 47 of 219 GCs, which are saturated in some filters,
are also excluded in our sample. At last 172 GCs are included here. 
Since our sample candidates in Bologna catalog are classified into class A and B by
authors' high level of confidence, and cross-checked by others (e.g. Barmby et al. 
2000), we consider that these objects are GCs, and we do not confirm them again.
In addition, we should emphasize that the numbering system of the Bologna catalog
is adopted in this paper.

\subsection{Observations and Data Reduction}

The large field multicolor observations of M31 were obtained in the BATC 
photometric system, in which a $60/90$ cm f/3 Schmidt telescope is used. 
This telescope is located at Xinglong Station of National Astronomical 
Observatories. A Ford Aerospace 
2048$\times$2048 CCD camera, with 15 $\mu$m pixel size is mounted at the 
Schmidt focus, giving a CCD field of view of 
$58^{\prime}$ $\times $ $58^{\prime}$ with a pixel size of 
$1\arcsec{\mbox{}\hspace{-0.15cm}.} 7$.

The multiband BATC filter system comprises 15 intermediate-band
filters, covering the full optical wavelength range from 3000 to
10000{\AA}. The filters were designed specifically to avoid
contamination from the brightest and most variable night sky emission
lines. Details of the Schmidt Telescope, the CCD camera, the data
acquisition system, and the definition of the BATC filter passbands
can be found in previous publications (Fan et al. 1996; Zheng et al.
1999). The observations of M31 were carried out from September 15, 1995,
through December 16, 1999, with a total exposure time of about 37 hr
20 min, while the CCD images were accumulated in 13 of the BATC filters.
The dome flats were obtained using 
a diffusor plate in front of the Schmidt corrector plate. For flux
calibration the Oke-Gunn primary standard stars HD~19445, HD~84937,
BD~+26$^\circ$2606, and BD~+17$^\circ$4708 were observed under
photometric conditions (see Yan et al. 2000, Zhou et al. 2001 for
details). The parameters of the filters, and the basic statistics
of the observations are given in Table 1.    

Using standard procedures, the data were reduced by automatic reduction
software: PIPELINE I, which include bias subtraction and flat-fielding
of the CCD images. This software was developed for the BATC multicolor
sky survey (see Ma et al. 2001, 2002a for a detail). The absolute flux of
images was calibrated using observations of standard stars. Fluxes,
as observed through the BATC filters for the Oke-Gunn stars, were derived
by convolving the SEDs of these stars with the measured BATC filter transmission
functions (Fan et al. 1996). In Table 1, {\it Column 6}
gives the zero point errors in magnitude for the standard stars
through each filter. The formal errors obtained for these stars in
the 13 BATC filters used are $\la 0.02$ mag, which implies that we can
define photometrically the BATC system to an accuracy of better than
0.02 mag. 

\setcounter{table}{0} 
\begin{table}[ht]
\caption[]{Parameters of the BATC Filters
and Statistics of Observations}
\vspace {0.5cm}
\begin{tabular}{cccccc}
\hline
\hline
 No. & Name& cw\tablenotemark{a}~~(\AA)& Exp. (hr)&  N.img\tablenotemark{b}
 & rms\tablenotemark{c} \\
\hline
1  & BATC03& 4210   & 01:00& 03 &0.015\\
2  & BATC04& 4546   & 05:30& 17 &0.009\\
3  & BATC05& 4872   & 03:30& 11 &0.015\\
4  & BATC06& 5250   & 02:20& 12 &0.006\\
5  & BATC07& 5785   & 02:15& 07 &0.003\\
6  & BATC08& 6075   & 01:40& 05 &0.003\\
7  & BATC09& 6710   & 00:45& 03 &0.003\\
8  & BATC10& 7010   & 03:00& 12 &0.008\\
9  & BATC11& 7530   & 02:00& 06 &0.004\\
10 & BATC12& 8000   & 04:00& 12 &0.003\\
11 & BATC13& 8510   & 01:30& 05 &0.004\\
12 & BATC14& 9170   & 05:50& 18 &0.003\\
13 & BATC15& 9720   & 04:00& 12 &0.009\\      
\hline
\end{tabular}\\
\tablenotetext{a}{Central wavelength for each BATC filter}
\tablenotetext{b}{Image numbers for each BATC filter}
\tablenotetext{c}{Zero point error, in magnitude, for each filter
as obtained from the standard stars}
\end{table}

\subsection{Integrated Photometry}

To obtain the magnitude of a given GC, the PHOT routine in DAOPHOT (Stetson 1987, 
1990) is used. For avoiding contamination from nearby objects, We adopt a small 
aperture diameter of $10\arcsec{\mbox{}\hspace{-0.15cm}.} 2$, which corresponds 
to a diameter of 6 pixels in Ford CCDs. Aperture corrections are computed 
using isolated stars. Finally we obtained the SEDs in 13 BATC filters for 
172 GCs, which are listed in Table 2. The table contains the following 
information: {\it Column 1} is the cluster number taken from the Bologna 
catalog (Battistini et al. 1987). {\it Column} 2 to {\it Column} 14 present the 
magnitudes in the selected BATC bands, and on a second row for each GC in these 
columns we give the magnitude uncertainty for each band. The uncertainties are 
given by DAOPHOT. 

\subsection{Magnitudes in the {\it B} and {\it V} Bands, and $B-V$ Colors}

Using Landolt standards and the catalogs of Landolt (1983, 1992) and of
Galad\'\i-Enr\'\i quez et al. (2000), Zhou et al. (2002) derived the 
relationships between the BATC intermediate band system and the {\it UBVRI} 
broad-band system. These relationships are given in equations (1) and (2) as:
\beq
m_B=m_{04}+(0.2218\pm0.033)(m_{03}-m_{05})+0.0741\pm0.033,
\eeq
\beq
m_V=m_{07}+(0.3233\pm0.019)(m_{06}-m_{08})+0.0590\pm0.010.
\eeq
Using equations (1) and (2) we transformed the magnitudes of the 172
GCs in the BATC03, BATC04 and BATC05 bands into {\it B} band magnitudes,
and also derived {\it V} band magnitudes from those in BATC06, BATC07 and BATC08.

Histograms of the computed {\it V} magnitudes and $B-V$ colors, and color-magnitude 
diagram, {\it V} against $B-V$, are shown in Figure 1. The figure is in good 
agreement with the corresponding plots of class A and B GCs given by Battistini et 
al. (1987). From the left panel of Figure 1 we can see that most of our sample GCs 
are brighter than $V=18$. This relates to the original catalog, because {\it V} 
magnitudes of most class A and B GCs in Bologna catalog are brighter than 18 mag. 
The right panel of Figure 1 shows that $B-V$ colors of most GCs are redder than 0.6, 
which is reasonable. Actually, the GCs are generally the objects with $B-V>0.6$ 
(Crampton et al. 1985) or $B-V>0.55$ (Barmby et al. 2000). Iye \& Richter (1985) 
even considered $B-V<0.6$ as one of their criteria of rejecting GC candidates.

We also present the comparisons of the BATC photometry with previously published 
measurements (Battistini et al. 1987; Barmby et al. 2000) in Figures 2 and 3. These 
figures do not present all GCs of our sample, since {\it V} magnitudes and $B-V$ 
colors of some GCs were not given by Battistini et al. (1987) and Barmby et al. 
(2000). Figure 2 plots the comparison for 157 objects, while Figure 3 for 141 
objects. From these figures, it can be seen that there are good agreements 
between the BATC photometry and the other two photometric measurements. 
In Figure 2, the differences of {\it V} magnitudes for several objects are 
abnormally big, such as clusters 118, 127, 131, 138, and 186. 
The reason may be that clusters 118, 127, 131 and 138 lie in the center region of 
the parent galaxy, so we can not subtract the background very well. However, 
the BATC {\it V} magnitude of cluster 186 is in good agreement with the value presented by
Barmby et al. (2000). The mean differences of {\it V} and $B-V$ between the BATC 
photometry and Battistini et al. (the BATC photometric values minus the values of 
Battistini et al.) are $<\Delta V>=-0.083\pm0.069$ and $<\Delta (B-V)>=-0.128\pm0.096$ 
(excluding objects mentioned above), and between the BATC photometry and Barmby et 
al. (2000) are $<\Delta V>=0.063\pm0.065$ and $<\Delta (B-V)>=0.067\pm0.090$, 
respectively. The uncertainties in {\it B} (BATC) and {\it V} (BATC) were determined 
linearly, i.e. $\sigma_B=\sigma_{04}+0.2218(\sigma_{03}+\sigma_{05})$ and 
$\sigma_V=\sigma_{07}+0.3233(\sigma_{06}+\sigma_{08})$, to reflect photometric errors 
in the six filter bands. For the colors, we calculated their errors by 
$\sigma_{B-V}=\sqrt{\sigma_{B}^2+\sigma_{V}^2}$.
From these two figures, we can see that, our results are in
better agreement with Barmby et al. (2000) than with Battistini et al. (1987). 

\section{DATABASES OF SIMPLE STELLAR POPULATIONS}

A simple stellar population (SSP) is defined as a single generation of coeval 
stars with fixed parameters such as metal abundance, initial mass function, etc. 
(Buzzoni 1997). In evolutionary synthesis models, they are modeled by a collection 
of evolutionary tracks of stars with different masses and initial chemical 
components, and a set of stellar spectra at different evolutionary stages. 
Because SSPs are the basic building elements for synthetic spectra of galaxies, 
we can infer the formation and evolution of the parent galaxies from them 
(Jablonka et al. 1996). 
Since Tinsley (1972) and Searle et al. (1973) did the pioneering work in 
evolutionary population synthesis, this method has become a standard technique 
to study the stellar populations of galaxies. A broad variety of empirical and 
theoretical database has been built up, and a comprehensive library of models 
has been compiled (Leitherer et al. 1996). Widely used models are from GISSEL96 
(Charlot \& Bruzual 1991; Bruzual \& Charlot 1993; Bruzual \& Charlot 1996, 
unpublished), the Padova and Geneva group (e.g. Schaerer \& de Koter 1997; Schaerer 
\& Vacca 1998; Bressan et al. 1996; Chiosi et al. 1998), PEGASE (Fioc \& 
Rocca-Volmerange 1997) and STARBURST99 (Leitherer et al. 1999). 

In present paper, we will use the SSPs of Galaxy Isochrone Synthesis Spectra 
Evolution Library (hereafter GSSP; Bruzual \& Charlot 1996, unpublished) 
to estimate the ages of the sample GCs, since they are simple and well explored. 

\subsection{Spectral Energy Distribution of GSSPs}

GSSPs method, which is based on a model of stellar population synthesis developed 
by Charlot \& Bruzual (1991), can be used to 
determine the distribution of stars in the theoretical color-magnitude diagram 
for any stellar system. The updated GSSPs synthesis model (Bruzual \& Charlot 
1996, unpublished), which upgraded from the Bruzual \& Charlot (1993) version, 
provides the evolution of the spectra photometric properties for a wider range 
of stellar metallicities, with $Z$ = 0.0004, 0.004, 0.008, 0.02, 0.05, and 0.1. 
The chemical parameters follow the helium-to-metal enrichment law $dY/dZ=2.5$, 
and the initial mass function obeys the Salpeter (1955) law with $\alpha=2.35$ 
(Leitherer et al. 1996).

\subsection{Integrated Colors of GSSPs}

To obtain the age, metallicity, and interstellar-medium reddening distribution 
for M81, Kong et al. (2000) found the best match between the intrinsic colors 
and the predictions of GSSP for each cell of M81. To estimate the ages for 
the sample clusters in this paper, we follow the method of Kong et al. (2000). 
Since the observational data are integrated luminosity, as Kong et al. (2000) and 
Ma et al. (2002b) 
did, we convolve the SED of GSSP with the BATC filter profiles to obtain the optical 
and near-infrared integrated luminosity for comparisons. The integrated luminosity 
$L_{\lambda_i}(t,Z)$ in the $i$th BATC filter can be calculated with
\beq
L_{\lambda_i}(t,Z) =\frac{\int
F_{\lambda}(t,Z)\varphi_i(\lambda)d\lambda} {\int
\varphi_i(\lambda)d\lambda},
\eeq
where $F_{\lambda}(t,Z)$ is the SED of the GSSP of metallicity $Z$ at age $t$, 
and $\varphi_i(\lambda)$ is the response function in the $i$th filter of the BATC 
filter system ($i=3, 4, \cdot\cdot\cdot, 15$), respectively.      
To avoid using parameters that are dependent on the distance, we calculate 
the integrated colors of a GSSP relative to the BATC filter BATC08 
($\lambda=6075${\AA}):
\beq
\label{color}
C_{\lambda_i}(t,Z)={L_{\lambda_i}(t,Z)}/{L_{6075}(t,Z)}.
\eeq         
Finally, using equations (3) and (4) we obtained the intermediate-band colors 
of a GSSP for six metallicities from $Z = 0.0004$ to $Z = 0.1$.

\section{REDDENING CORRECTION}

The SED of a stellar system will be affected by age, metallicity and reddening 
along the line of sight. Generally, older age, higher metallicity and larger 
reddening all lead to redder SEDs of stellar systems in the optical 
(Bressan, Chiosi, \& Tantalo 1996; Moll\`{a}, Ferrini, \& Diaz 1997), 
and these effects are difficult to separate (Calzetti 1997; Vazdekis 
et al. 1997; Origlia et al. 1999). In order to estimate the ages for 172 GCs, 
the intrinsic colors of these GCs should be obtained. The observed colors 
are mainly affected by two sources of reddening: the foreground extinction in 
the Milk Way and internal reddening in M31.

The Galactic reddening in the direction of M31 was estimated by many authors 
(e.g. van den Bergh 1969; McClure \& Racine 1969; Frogel, Persson, \& Cohen 1980), 
and the similar values of the foreground color excess, $E(B-V)$, were determined, 
such as $E(B-V)=0.08$ given by van den Bergh (1969), 0.11 given by McClure \& 
Racine (1969), and 0.08 given by Frogel, Persson, \& Cohen (1980). As Crampton 
et al. (1985) did, we use the value of 0.10 as the foreground color excess.

The reddening of GCs in M31 has been determined by several ways in previous studies. 
As is mentioned, Vete\u{s}nik (1962a) compiled a comprehensive catalog of 257 GC 
candidates, and derived color excesses for the candidates 
that were considered to be most probably the GCs (Vete\u{s}nik 1962b). 
To obtain true colors of the clusters, Vete\u{s}nik (1962b) 
calculated the average true color index of 36 GCs beyond the body of M31, and 
considered the result, $B-V=0.83$ mag, as the uniform value of true color for all 
GCs in M31. Actually, this implicated that these clusters were only affected 
by the foreground Galactic extinction. Later, many authors used this 
assumption of a single intrinsic color for all GCs in M31 (Bajaja \& 
Gergely 1977; Iye \& Richter 1985).

Using the slope parameter, S, Crampton et al. (1985) calculated intrinsic 
colors for individual GCs. S, defined by Hartwick (1968), was proved to be a 
good indicator of intrinsic color and metallicity. From the derived relationship 
between $(B-V)_0$ and S, Crampton et al. (1985) gave intrinsic colors and 
color excesses for most candidates in their catalog.

A comprehensive list of colors and metallicities for the M31 GCs was given 
by Barmby et al. (2000). In order to determine the cluster reddening, they set 
two reasonable 
assumptions that both the extinction law and the GC intrinsic color were the 
same. By using the color-metallicity relationships and the relationships 
between colors and reddening-free parameters, two basic methods were used 
independently to obtain the intrinsic colors. Finally, for each GC, the two 
results were combined by their own weight.

In present paper, we mainly use color excesses given by Barmby et al. (2000) 
for reddening correction. Since our sample contains a total of 172 GCs, and 
values of $E(B-V)$ for only 152 of 172 objects were derived by Barmby et al. 
(2000), there remains 20 objects undetermined. We note that values of S, the 
slope parameter, for 9 of these 20 GCs were given by Crampton et al. (1985), 
the intrinsic colors for these 9 GCs can be derived from the equation given by 
Crampton et al. (1985):
\beq  
(B-V)_0=0.066S-0.17(B-V)+0.32,
\eeq 
where the values of S are taken from Crampton et al. (1985) and values of 
$B-V$ comes from Barmby et al. (2000). At last there are still 11 GCs whose 
color excesses are not determined. For these 11 objects, we assume that they 
are only affected by the foreground Galactic extinction and their color excesses 
are the foreground color excess, i.e., $E(B-V)=0.10$. 
In addition, we adopted the extinction curve presented by Zombeck (1990).
An extinction correction $A_{\lambda}=R_{\lambda}E(B-V)$ was
applied, here $R_{\lambda}$ is obtained by interpolating
using the data of Zombeck (1990).

\section{AGE ESTIMATES}

After the photometric measurements are dereddened, intrinsic colors for each 
GC depend on two parameters, age and metallicity, since we model the stellar 
populations of the GCs by SSPs. In this section we will determine the ages
and best-fitted model of metallicity
for our sample GCs simultaneously by the least-square method. 
The age and best-fitted models of metallicity are found by minimizing the 
difference between the intrinsic colors of the sample GCs and integrated colors 
of GSSP:
\beq
R^2(n,t,Z)=\frac{\sum_{i=3}^{15}{[C_{\lambda_i}^{\rm intr}(n)-C_{\lambda_i}^ 
{\rm ssp}(t, Z)]^2}/\sigma_{i}^{2}} {{\sum_{i=3}^{15}}{1/\sigma_{i}^{2}}}, 
\eeq    
where $C_{\lambda_i}^{\rm ssp}(t, Z)$ is the integrated color in the $i$th 
filter of a SSP at age $t$ in the model of metallicity $Z$, and 
$C_{\lambda_i}^{\rm intr}(n)$ presents the intrinsic integrated color in the 
same filter of the $n$th GC. The differences are weighted by $1/\sigma_{i}^{2}$, 
where the ${\sigma_i}$'s are observational uncertainties of the passbands. 
The M31 GCs generally have a metal 
abundance , $[Fe/H]$, lower than 0.0 (Barmby et al. 2000), which corresponds to 
$Z = 0.0169$ (Leitherer et al. 1996), so we only select the models of three 
metallicity, 0.0004, 0.004 and 0.02 of GSSP. 

Figure 4 shows the fit of the integrated color of a SSP ($Z$ = 0.0004, 0.004 
and 0.02) with the intrinsic color for 20 GCs selected from the 172 GCs (the 
first 20 GCs in Table 2). In Figure 4, filled circle represents the intrinsic 
integrated color of a GC, and the thick line represents the best fit of the 
integrated color of a SSP of GSSP. From Figure 4, we see that SEDs of GCs are 
fitted very well by the best-fitted SSP of GSSP model. Table 3 presents ages 
and the best-fitted 
models of metallicities for all the 172 GCs. The uncertainties in the age 
estimates arising from photometric uncertainties are 0.2 or so, i.e, 
$\rm{age}\pm 0.2\times\rm{age}~[\log \rm{yr}]$. In addition, we noted that 
clusters 127 and 225 have strong emission lines in the filter of BATC09, so we 
did not use the color of this filter in the process of fitting.  

We should emphasize that, in this study, we estimate the ages of our sample clusters
by comparing the photometry of each object with models for different values of metallicity
as Chandar, Bianchi, \& Ford (1999b, 1999c, 2001) did. Recently, using the
similar technique, Ma et al. (2002a) estimated ages of 10 halo GCs in M33
with four models of metallicities ($Z$ = 0.0004, 0.004, 0.008, and 0.02).
Here we use three metallicity models to estimate
ages for our sample GCs. In each model, the ages of SSPs are
from 0 to 20 Gyrs.
We should also emphasize that, for very old globular clusters,
the age/metallicity degeneracy becomes pronounced. In this case,
we only mean that in some model of metallicity, the intrinsic integrated color
of a GC can do the best fit with the integrated color of a SSP at some age.

From Table 3, we see that most GCs are old objects, except clusters 133 and 362. 
Cluster 133 appears to be very young, and its reddening-corrected SED significantly 
differs from others'. This case can be explained by the high value
of $E(B-V)$. Since $E(B-V)$ of cluster 133 was not given by Barmby et al. (2000),
we calculated its value using S (see Sec. 4). According to equation (5), $E(B-V)$
is determined as follows:
\beq
E(B-V)=1.17(B-V)-0.066S-0.32,
\eeq
where $B-V$ comes from Barmby et al. (2000), and S from Crampton et al. (1985),
as has been described in Sec. 4. From equation (7), we can see that a high value of 
$B-V$ and a small value of S all lead to high $E(B-V)$. Besides, we should note that
the value of $B-V$ of cluster 133 given by Barmby et al. (2000) is 0.93, while the 
corresponding values given by Battistini et al. (2000) and the BATC photometry are 
0.66 and 0.26 respectively. The value of S given by Crampton et al.
(1985) is very low, $-4$, while values of most S are more than 0. Due to these
two reasons, or maybe one of the two reasons, cluster 133 appears very young. 
Another ``young'' cluster is cluster 362, whose $E(B-V)$ given by Barmby et al. 
(2000) is 0.42, but the uncertainty of $E(B-V)$ is 0.45. Perhaps due to the big 
uncertainty of $E(B-V)$, cluster 362 also appears very young. In the next analysis, 
we do not include these two GCs.

Figure 5 plots histograms of GC ages for three models of different metallicities,
$Z$ = 0.02, 0.004, and 0.0004, and for all GCs, except clusters 133 and 362.  
From this figure, we can see that almost all these GCs have ages more than 
$10^{9}$ years, and most of them are around $10^{10}$ years old. In 
the separated histograms of three models, only a few GCs are included in $Z = 0.02$
model, while most GCs are included in $Z=0.004$ and $Z = 0.0004$ models. 
We can also see that ages in $Z=0.004$ and $Z=0.0004$ models have apparently 
different distributions. The peak of age in $Z=0.004$ model is at 
about 6 Gyr, while the peak in $Z=0.0004$ model is about 19 Gyr. 
This means, in general, GCs in $Z=0.0004$ model are older than those in 
$Z=0.004$ model.

Barmby \& Huchra (2000) compared simple stellar population colors of three
population synthesis models to the intrinsic colors of Galactic and M31 GCs in
$UBVRIJHK$ colors. They found that higher metallicity cluster colors are best fit
by the younger models, and lower metallicity cluster colors are best fit by the
older models. Our results in this paper are in agreement with Barmby \& Huchra (2000).

It is well known that the abundance distributions of GCs in many galaxies are
bimodal, including M31 GC system (e.g. Barmby et al. 2000; Perrett et al. 2002).
Forbes, Brodie, \& Grillmair (1997) found that the metal-rich GCs in elliptical 
and cD galaxies are closely coupled to their parent galaxies, but the metal-poor 
GCs are largely independent of the galaxies. They concluded that the metal-poor 
GCs are formed during the beginning of galaxy formation, and the metal-rich GCs 
are formed at a later stage than metal-poor GCs. If so, this may imply 
that the age distribution of GCs should be bimodal or multi-modal. 
Due to our incomplete sample, from Figure 5 we can not confirm whether or not 
the age distribution is bimodal or multi-modal, however, we can say that 
the age distribution of GCs in M31 is not monomodel.

The presence of groups of clusters in space and
in age is interesting. We show the GC age
as a function of galactocentric distance $R$ in Figure 6. 
The galactocentric distance $R$ for M31 GCs is derived using the distance 
modulus of 24.47 (Holland 1998; Stanek \& Garnavich 1998), the inclination angle 
of 77 degrees (Williams \& Hodge 2001), and the position angle of 38 degrees 
(Williams \& Hodge 2001). Figure 6 shows that, there
exits a few groups of clusters which are nearby in space and in age.
This result tells us that some nearby GCs formed simultaneously.
We also plot age versus reddening-corrected apparent magnitude in Figure 7,
no trend is obvious.

\section{SUMMARY} 

In the present paper, we for the first time obtained the SEDs for 172 GCs of 
M31 in 13 intermediate band filters, with the BATC 60/90 cm Schmidt telescope. 
The main results and conclusions are summarized as follows:

1. Using the images obtained with the BATC Multicolor Sky Survey Telescope, we obtained 
SEDs for 172 GCs of M31 selected from the Bologna catalog, in 13 intermediate-band 
filters covering a range of wavelength from 3800 to 10000{\AA}. We also gave 
identification charts for 219 class A or B GCs of Bologna catalog in our CCD 
field.

2. Using the relationship between the BATC intermediate-band system and the 
{\it UBVRI} broad-band system, we derived the magnitudes in the {\it B} and {\it V} 
bands. The computed {\it V} and $B-V$ are in agreement with previous measurements.

3. By comparing the photometry of each GC with theoretical stellar population 
synthesis models, we estimated ages of the sample GCs for different metallicities. 
The results show that nearly all the GCs have ages more than $10^{9}$ years, and
GCs in the metal-poor model are generally older than ones in the metal-rich model.


\acknowledgments 
We would like to thank the anonymous referee for his/her
insightful comments and suggestions that improved this paper.
We are grateful to P. Barmby and J. P. Huchra for providing us values of 
$E(B-V)$ for GCs in M31 that they derived. 
We would like to thank the Padova group for providing us with a set of
theoretical isochrones and SSPs. We are also indebted to G. Bruzual and
S. Charlot for sending us their latest calculations of SSPs and
for explanations of their code.
The work is supported partly by the National Sciences
Foundation under  the contract  No.19833020 and No.19503003.
The BATC Survey is supported by the
Chinese Academy of Sciences, the Chinese National Natural Science
Foundation and the Chinese State Committee of Sciences and
Technology.
The project is also supported in part
by the National Science Foundation (grant INT 93-01805) and
by Arizona State University, the University of Arizona and Western
Connecticut State University.


\clearpage
{\small
\setcounter{table}{1}
\begin{table}[ht]
\caption{SEDs of 172 GCs in M31}
\vspace {0.3cm}
\begin{tabular}{cccccccccccccc}
\hline
\hline
 No. & 03  &  04 &  05 &  06 &  07 &  08 &  09 &  10 &  11 &  12 &  13 &  14 &  15\\
(1)    & (2) & (3) & (4) & (5) & (6) & (7) & (8) & (9) & (10) & (11) & (12) & (13) & (14)\\
\hline          
   326 &  16.18 &  16.16 &  16.01 &  16.00 &  15.78 &  15.81 &  15.77 &  15.69 &  15.64 &  15.51 &  15.47 &  15.43 & 15.44 \\        &  0.175 &  0.173 &  0.185 &  0.188 &  0.186 &  0.197 &  0.198 &  0.202 &  0.206 &  0.211 &  0.220 &  0.212 & 0.219 \\
   328 &  17.43 &  17.14 &  16.92 &  16.84 &  16.50 &  16.49 &  16.42 &  16.31 &  16.21 &  16.03 &  15.96 &  15.96 & 16.01 \\        &  0.106 &  0.162 &  0.163 &  0.162 &  0.153 &  0.159 &  0.159 &  0.158 &  0.165 &  0.159 &  0.163 &  0.161 & 0.177 \\
   332 &  19.27 &  18.90 &  18.74 &  18.91 &  18.59 &  18.58 &  18.51 &  18.68 &  18.22 &  17.96 &  17.67 &  18.16 & 17.67 \\        &  0.250 &  0.225 &  0.262 &  0.321 &  0.337 &  0.356 &  0.375 &  0.488 &  0.369 &  0.340 &  0.301 &  0.463 & 0.311 \\
\end{tabular}
\end{table}
}             

\clearpage
{\small
\setcounter{table}{1}
\begin{table}[ht]
\caption{Continued}
\vspace {0.3cm}
\begin{tabular}{cccccccccccccc}
\hline
\hline
 No. & 03  &  04 &  05 &  06 &  07 &  08 &  09 &  10 &  11 &  12 &  13 &
14 &  15\\
(1)    & (2) & (3) & (4) & (5) & (6) & (7) & (8) & (9) & (10) & (11) & (12) & (13) & (14)\\
\hline             
    33 &  19.03 &  18.80 &  18.45 &  18.18 &  17.89 &  17.70 &  17.52 &  17.35 &  17.45 &  17.35 &  17.29 &  17.16 & 17.05 \\        &  0.071 &  0.074 &  0.067 &  0.056 &  0.056 &  0.045 &  0.044 &  0.043 &  0.052 &  0.050 &  0.064 &  0.065 & 0.075 \\
    34 &  16.46 &  16.03 &  15.77 &  15.59 &  15.28 &  15.23 &  14.98 &  14.87 &  14.76 &  14.71 &  14.74 &  14.49 & 14.50 \\        &  0.006 &  0.004 &  0.003 &  0.003 &  0.004 &  0.004 &  0.004 &  0.004 &  0.004 &  0.004 &  0.008 &  0.005 & 0.007 \\
    36 &  18.38 &  17.96 &  17.64 &  17.45 &  17.09 &  16.97 &  16.79 &  16.70 &  16.53 &  16.45 &  16.34 &  16.20 & 16.20 \\        &  0.022 &  0.014 &  0.010 &  0.011 &  0.013 &  0.010 &  0.012 &  0.012 &  0.016 &  0.012 &  0.030 &  0.021 & 0.030 \\
\end{tabular}
\end{table}
}     

\clearpage
{\small
\setcounter{table}{1}
\begin{table}[ht]
\caption{Continued}
\vspace {0.3cm}
\begin{tabular}{cccccccccccccc}
\hline
\hline
 No. & 03  &  04 &  05 &  06 &  07 &  08 &  09 &  10 &  11 &  12 &  13 &
14 &  15\\
(1)    & (2) & (3) & (4) & (5) & (6) & (7) & (8) & (9) & (10) & (11) & (12) & (13) & (14)\\
\hline              
    63 &  17.06 &  16.51 &  16.08 &  15.85 &  15.36 &  15.26 &  14.99 &  14.84 &  14.64 &  14.55 &  14.46 &  14.19 & 14.16 \\        &  0.011 &  0.007 &  0.005 &  0.005 &  0.004 &  0.004 &  0.004 &  0.004 &  0.004 &  0.004 &  0.006 &  0.004 & 0.006 \\
    64 &  17.01 &  16.72 &  16.45 &  16.29 &  16.05 &  15.99 &  15.82 &  15.73 &  15.62 &  15.56 &  15.49 &  15.39 & 15.44 \\        &  0.019 &  0.021 &  0.021 &  0.020 &  0.022 &  0.021 &  0.021 &  0.023 &  0.025 &  0.028 &  0.031 &  0.031 & 0.036 \\
    67 &  17.90 &  17.60 &  17.37 &  17.24 &  17.02 &  16.99 &  16.81 &  16.73 &  16.69 &  16.69 &  16.51 &  16.59 & 16.71 \\        &  0.020 &  0.021 &  0.021 &  0.022 &  0.026 &  0.026 &  0.032 &  0.031 &  0.042 &  0.043 &  0.052 &  0.057 & 0.078 \\
\end{tabular}
\end{table}
}       

\clearpage
{\small
\setcounter{table}{1}
\begin{table}[ht]
\caption{Continued}
\vspace {0.3cm}
\begin{tabular}{cccccccccccccc}
\hline
\hline
 No. & 03  &  04 &  05 &  06 &  07 &  08 &  09 &  10 &  11 &  12 &  13 &
14 &  15\\
(1)    & (2) & (3) & (4) & (5) & (6) & (7) & (8) & (9) & (10) & (11) & (12) & (13) & (14)\\
\hline            
    88 &  16.64 &  16.26 &  15.86 &  15.60 &  15.16 &  15.05 &  14.80 &  14.66 &  14.52 &  14.36 &  14.26 &  14.09 & 14.06 \\        &  0.008 &  0.006 &  0.004 &  0.004 &  0.004 &  0.003 &  0.003 &  0.003 &  0.003 &  0.003 &  0.005 &  0.004 & 0.005 \\
    90 &  19.19 &  18.53 &  18.33 &  18.25 &  18.03 &  18.05 &  17.81 &  17.80 &  17.70 &  17.87 &  17.83 &  17.44 & 17.51 \\        &  0.133 &  0.091 &  0.103 &  0.106 &  0.102 &  0.106 &  0.102 &  0.113 &  0.126 &  0.169 &  0.193 &  0.147 & 0.183 \\
    91 &  17.82 &  17.61 &  17.49 &  17.47 &  17.27 &  17.28 &  17.25 &  17.18 &  17.21 &  17.15 &  17.04 &  17.06 & 17.23 \\        &  0.039 &  0.041 &  0.049 &  0.053 &  0.057 &  0.061 &  0.069 &  0.073 &  0.090 &  0.094 &  0.111 &  0.129 & 0.166 \\
\end{tabular}
\end{table}
}  

\clearpage
{\small
\setcounter{table}{1}
\begin{table}[ht]
\caption{Continued}
\vspace {0.3cm}
\begin{tabular}{cccccccccccccc}
\hline
\hline
 No. & 03  &  04 &  05 &  06 &  07 &  08 &  09 &  10 &  11 &  12 &  13 &
14 &  15\\
(1)    & (2) & (3) & (4) & (5) & (6) & (7) & (8) & (9) & (10) & (11) & (12) & (13) & (14)\\
\hline      
   109 &  17.46 &  17.00 &  16.66 &  16.51 &  16.15 &  16.05 &  15.88 &  15.74 &  15.55 &  15.50 &  15.30 &  15.14 & 15.09 \\        &  0.046 &  0.049 &  0.044 &  0.046 &  0.052 &  0.049 &  0.049 &  0.050 &  0.053 &  0.057 &  0.055 &  0.053 & 0.056 \\
   111 &  17.51 &  17.23 &  17.00 &  16.84 &  16.60 &  16.57 &  16.36 &  16.28 &  16.25 &  16.22 &  16.15 &  16.07 & 16.06 \\        &  0.017 &  0.014 &  0.012 &  0.011 &  0.013 &  0.012 &  0.012 &  0.013 &  0.016 &  0.014 &  0.026 &  0.024 & 0.031 \\
   110 &  16.08 &  15.69 &  15.41 &  15.25 &  14.97 &  14.92 &  14.71 &  14.60 &  14.50 &  14.44 &  14.36 &  14.23 & 14.23 \\        &  0.005 &  0.004 &  0.003 &  0.004 &  0.004 &  0.003 &  0.004 &  0.004 &  0.004 &  0.004 &  0.007 &  0.006 & 0.008 \\
\end{tabular}
\end{table}
}   

\clearpage
{\small
\setcounter{table}{1}
\begin{table}[ht]
\caption{Continued}
\vspace {0.3cm}
\begin{tabular}{cccccccccccccc}
\hline
\hline
 No. & 03  &  04 &  05 &  06 &  07 &  08 &  09 &  10 &  11 &  12 &  13 &
14 &  15\\
(1)    & (2) & (3) & (4) & (5) & (6) & (7) & (8) & (9) & (10) & (11) & (12) & (13) & (14)\\
\hline                         
   133 &  17.47 &  17.47 &  17.33 &  17.30 &  17.23 &  17.21 &  17.11 &  16.97 &  16.74 &  16.71 &  16.60 &  16.57 & 16.59 \\        &  0.100 &  0.123 &  0.141 &  0.142 &  0.167 &  0.169 &  0.165 &  0.159 &  0.138 &  0.152 &  0.166 &  0.153 & 0.163 \\
   134 &  17.14 &  16.71 &  16.49 &  16.29 &  15.98 &  15.90 &  15.74 &  15.60 &  15.51 &  15.42 &  15.23 &  15.18 & 15.32 \\        &  0.187 &  0.190 &  0.219 &  0.200 &  0.220 &  0.219 &  0.221 &  0.221 &  0.247 &  0.270 &  0.265 &  0.261 & 0.310 \\
   135 &  16.95 &  16.60 &  16.26 &  16.08 &  15.73 &  15.64 &  15.46 &  15.35 &  15.24 &  15.15 &  15.05 &  14.94 & 14.91 \\        &  0.008 &  0.007 &  0.005 &  0.006 &  0.006 &  0.005 &  0.006 &  0.006 &  0.007 &  0.007 &  0.010 &  0.009 & 0.012 \\
\end{tabular}
\end{table}
}  

\clearpage
{\small
\setcounter{table}{1}
\begin{table}[ht]
\caption{Continued}
\vspace {0.3cm}
\begin{tabular}{cccccccccccccc}
\hline
\hline
 No. & 03  &  04 &  05 &  06 &  07 &  08 &  09 &  10 &  11 &  12 &  13 &
14 &  15\\
(1)    & (2) & (3) & (4) & (5) & (6) & (7) & (8) & (9) & (10) & (11) & (12) & (13) & (14)\\
\hline                
   155 &  19.00 &  18.53 &  18.22 &  18.00 &  17.70 &  17.65 &  17.40 &  17.27 &  17.12 &  17.08 &  17.01 &  16.69 & 16.67 \\        &  0.049 &  0.034 &  0.025 &  0.023 &  0.025 &  0.019 &  0.023 &  0.022 &  0.024 &  0.023 &  0.042 &  0.027 & 0.055 \\
   157 &  18.23 &  17.98 &  17.74 &  17.60 &  17.45 &  17.40 &  17.28 &  17.21 &  17.28 &  17.16 &  17.04 &  17.01 & 17.08 \\        &  0.054 &  0.056 &  0.060 &  0.058 &  0.075 &  0.068 &  0.074 &  0.078 &  0.108 &  0.105 &  0.118 &  0.105 & 0.130 \\
   158 &  15.59 &  15.19 &  14.90 &  14.71 &  14.44 &  14.39 &  14.19 &  14.08 &  14.02 &  13.95 &  13.84 &  13.74 & 13.75 \\        &  0.004 &  0.003 &  0.002 &  0.002 &  0.002 &  0.002 &  0.002 &  0.002 &  0.002 &  0.002 &  0.004 &  0.003 & 0.005 \\
\end{tabular}
\end{table}
}  

\clearpage
{\small
\setcounter{table}{1}
\begin{table}[ht]
\caption{Continued}
\vspace {0.3cm}
\begin{tabular}{cccccccccccccc}
\hline
\hline
 No. & 03  &  04 &  05 &  06 &  07 &  08 &  09 &  10 &  11 &  12 &  13 &
14 &  15\\
(1)    & (2) & (3) & (4) & (5) & (6) & (7) & (8) & (9) & (10) & (11) & (12) & (13) & (14)\\
\hline       
   178 &  15.82 &  15.52 &  15.21 &  15.09 &  14.82 &  14.77 &  14.63 &  14.53 &  14.49 &  14.35 &  14.26 &  14.24 & 14.22 \\        &  0.006 &  0.005 &  0.005 &  0.005 &  0.005 &  0.005 &  0.005 &  0.006 &  0.007 &  0.007 &  0.009 &  0.009 & 0.011 \\
   179 &  16.23 &  15.87 &  15.55 &  15.42 &  15.14 &  15.09 &  14.95 &  14.83 &  14.75 &  14.65 &  14.56 &  14.50 & 14.53 \\        &  0.014 &  0.013 &  0.014 &  0.013 &  0.014 &  0.014 &  0.015 &  0.015 &  0.017 &  0.018 &  0.019 &  0.019 & 0.022 \\
   180 &  16.99 &  16.59 &  16.28 &  16.10 &  15.81 &  15.76 &  15.55 &  15.46 &  15.39 &  15.31 &  15.21 &  15.13 & 15.14 \\        &  0.008 &  0.006 &  0.005 &  0.005 &  0.006 &  0.004 &  0.005 &  0.005 &  0.007 &  0.007 &  0.012 &  0.010 & 0.015 \\
\end{tabular}
\end{table}
}  

\clearpage
{\small
\setcounter{table}{1}
\begin{table}[ht]
\caption{Continued}
\vspace {0.3cm}
\begin{tabular}{cccccccccccccc}
\hline
\hline
 No. & 03  &  04 &  05 &  06 &  07 &  08 &  09 &  10 &  11 &  12 &  13 &
14 &  15\\
(1)    & (2) & (3) & (4) & (5) & (6) & (7) & (8) & (9) & (10) & (11) & (12) & (13) & (14)\\
\hline    
   200 &  19.60 &  19.20 &  18.87 &  18.64 &  18.22 &  18.09 &  17.88 &  17.77 &  17.60 &  17.47 &  17.23 &  17.36 & 17.26 \\        &  0.217 &  0.186 &  0.170 &  0.139 &  0.124 &  0.110 &  0.097 &  0.098 &  0.101 &  0.096 &  0.093 &  0.130 & 0.145 \\
   201 &  16.85 &  16.56 &  16.28 &  16.14 &  15.87 &  15.82 &  15.64 &  15.55 &  15.46 &  15.37 &  15.14 &  15.25 & 15.25 \\        &  0.009 &  0.008 &  0.007 &  0.008 &  0.008 &  0.007 &  0.008 &  0.010 &  0.010 &  0.011 &  0.013 &  0.012 & 0.018 \\
   203 &  17.67 &  17.25 &  16.90 &  16.76 &  16.45 &  16.36 &  16.18 &  16.06 &  15.94 &  15.85 &  15.76 &  15.74 & 15.70 \\        &  0.019 &  0.015 &  0.013 &  0.013 &  0.012 &  0.011 &  0.012 &  0.011 &  0.015 &  0.012 &  0.021 &  0.019 & 0.024 \\
\end{tabular}
\end{table}
}  

\clearpage
{\small
\setcounter{table}{1}
\begin{table}[ht]
\caption{Continued}
\vspace {0.3cm}
\begin{tabular}{cccccccccccccc}
\hline
\hline
 No. & 03  &  04 &  05 &  06 &  07 &  08 &  09 &  10 &  11 &  12 &  13 &
14 &  15\\
(1)    & (2) & (3) & (4) & (5) & (6) & (7) & (8) & (9) & (10) & (11) & (12) & (13) & (14)\\
\hline                
   221 &  17.86 &  17.38 &  17.08 &  16.92 &  16.53 &  16.48 &  16.30 &  16.18 &  16.03 &  15.90 &  15.85 &  15.83 & 15.75 \\        &  0.016 &  0.012 &  0.011 &  0.010 &  0.010 &  0.009 &  0.010 &  0.011 &  0.014 &  0.011 &  0.019 &  0.019 & 0.025 \\
   222 &  18.20 &  17.82 &  17.64 &  17.50 &  17.28 &  17.26 &  17.08 &  17.00 &  16.86 &  16.82 &  16.77 &  16.58 & 16.51 \\        &  0.021 &  0.014 &  0.013 &  0.013 &  0.016 &  0.015 &  0.016 &  0.018 &  0.020 &  0.017 &  0.032 &  0.024 & 0.044 \\
   223 &  17.56 &  17.49 &  17.33 &  17.34 &  17.23 &  17.21 &  17.20 &  17.20 &  17.10 &  17.14 &  17.17 &  17.12 & 17.21 \\        &  0.022 &  0.026 &  0.029 &  0.026 &  0.033 &  0.032 &  0.039 &  0.040 &  0.049 &  0.047 &  0.073 &  0.074 & 0.118 \\
\hline
\end{tabular}
\end{table}
}


\clearpage
\setcounter{table}{2}
\begin{table}[htb]
\caption[]{Age Distribution of 172 GCs}
\vspace {0.3cm}
\begin{tabular}{ccc|ccc}
\hline
\hline
 No. & Metallicity ($Z$)& Age ([$\log$ yr]) & No. & Metallicity ($Z$)
 & Age ([$\log$ yr])\\
 (1) & (2) & (3) & (1) & (2) & (3) \\
\hline      
    326...... & 0.00400 &  8.806 &	     57...... & 0.00040 & 9.778 \\
    328...... & 0.00400 &  9.301 &	     59...... & 0.00400 & 9.740 \\
    332...... & 0.00400 &  9.057 &	     60...... & 0.00040 & 9.760 \\
    333...... & 0.02000 &  9.544 &	     61...... & 0.02000 & 9.362 \\
      9...... & 0.00040 & 10.068 &	     63...... & 0.00400 & 9.845 \\
     11...... & 0.00040 & 10.248 &	     64...... & 0.00040 & 10.279 \\
\end{tabular}
\end{table}  

\clearpage
\setcounter{table}{2}
\begin{table}[htb]
\caption[]{Continued}
\vspace {0.3cm}
\begin{tabular}{ccc|ccc}
\hline
\hline
 No. & Metallicity ($Z$)& Age ([$\log$ yr]) & No. & Metallicity ($Z$)
 & Age ([$\log$ yr])\\
 (1) & (2) & (3) & (1) & (2) & (3) \\
\hline      
    101...... & 0.02000 &  9.301 &	    141...... & 0.00400 & 9.439 \\
    103...... & 0.00400 & 10.130 &	    143...... & 0.00400 & 10.155 \\
    102...... & 0.00040 &  9.699 &	    144...... & 0.00400 & 10.021 \\
    104...... & 0.00040 &  9.574 &	    145...... & 0.02000 & 10.272 \\
    105...... & 0.00400 &  9.796 &	    146...... & 0.02000 & 9.279 \\
    106...... & 0.00400 &  9.796 &	    148...... & 0.00040 & 10.176 \\
\end{tabular}
\end{table}   

\clearpage
\setcounter{table}{2}
\begin{table}[htb]
\caption[]{Continued}
\vspace {0.3cm}
\begin{tabular}{ccc|ccc}
\hline
\hline
 No. & Metallicity ($Z$)& Age ([$\log$ yr]) & No. & Metallicity ($Z$)
 & Age ([$\log$ yr])\\
 (1) & (2) & (3) & (1) & (2) & (3) \\
\hline     
    180...... & 0.00040 & 10.041 &	    208...... & 0.00400 & 9.845 \\
    181...... & 0.00400 &  9.477 &	    209...... & 0.00400 & 9.677 \\
    182...... & 0.00400 &  9.720 &	    210...... & 0.00400 & 9.107 \\
    183...... & 0.00400 & 10.000 &	    211...... & 0.00040 & 10.301 \\
    185...... & 0.00400 &  9.760 &	    213...... & 0.00400 & 9.628 \\
    184...... & 0.00400 &  9.477 &	    214...... & 0.00040 & 10.283 \\
    186...... & 0.02000 & 10.301 &	    215...... & 0.00400 & 10.061 \\
    187...... & 0.00040 & 10.090 &	    216...... & 0.00040 & 8.957 \\
    188...... & 0.00400 & 10.283 &	    362...... & 0.00040 & 6.620 \\
    189...... & 0.02000 &  9.845 &	    217...... & 0.00400 & 9.720 \\
    190...... & 0.00400 &  9.813 &	    218...... & 0.00040 & 10.301 \\
    192...... & 0.00040 &  9.255 &	    220...... & 0.00040 & 10.079 \\
    194...... & 0.00040 & 10.301 &	    221...... & 0.00040 & 10.301 \\
    193...... & 0.00400 & 10.301 &	    222...... & 0.02000 & 9.107 \\
    197...... & 0.00040 &  9.057 &	    223...... & 0.00040 & 8.806 \\
    198...... & 0.00400 &  9.889 &	    224...... & 0.00040 & 9.860 \\
    200...... & 0.00040 &  9.875 &	    225...... & 0.00400 & 9.966 \\
    201...... & 0.00400 &  9.778 &	    228...... & 0.00400 & 10.090 \\
    203...... & 0.00040 & 10.297 &	    229...... & 0.00400 & 9.699 \\
    204...... & 0.00400 &  9.574 &	    231...... & 0.00400 & 9.929 \\
    205...... & 0.00400 &  9.740 &	    234...... & 0.00400 & 9.628 \\
    206...... & 0.00400 &  9.628 &	    235...... & 0.00400 & 9.778 \\
\hline
\end{tabular}
\end{table}


\setcounter{figure}{0}
\begin{figure}[ht]
\centerline{\epsfig{file=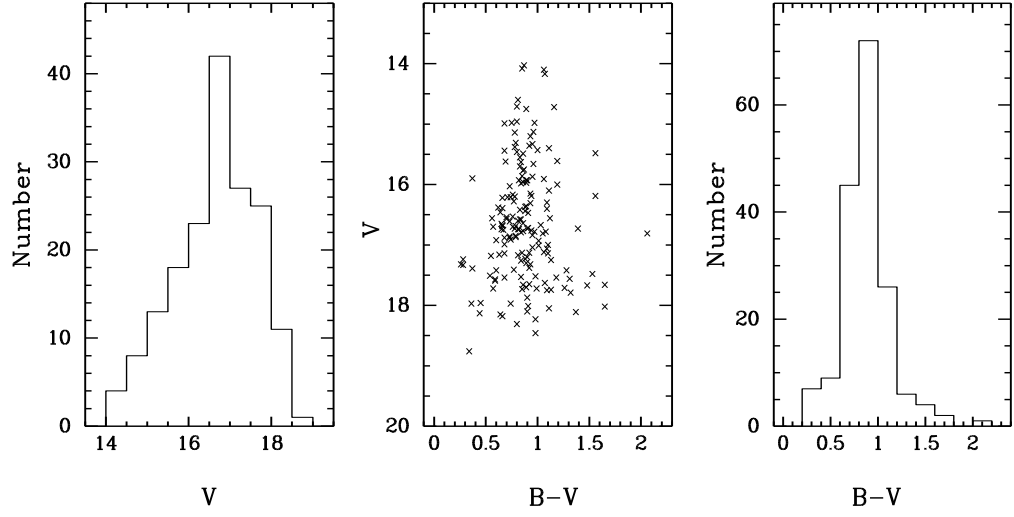,width=15.0cm,angle=-90}}
\vspace{-2.0cm}
\caption{Histograms of the computed V magnitudes and B-V colors, and 
color-magnitude diagram, for 172 GCs}
\end{figure}
 

\setcounter{figure}{1}
\begin{figure}[ht]
\centerline{\epsfig{file=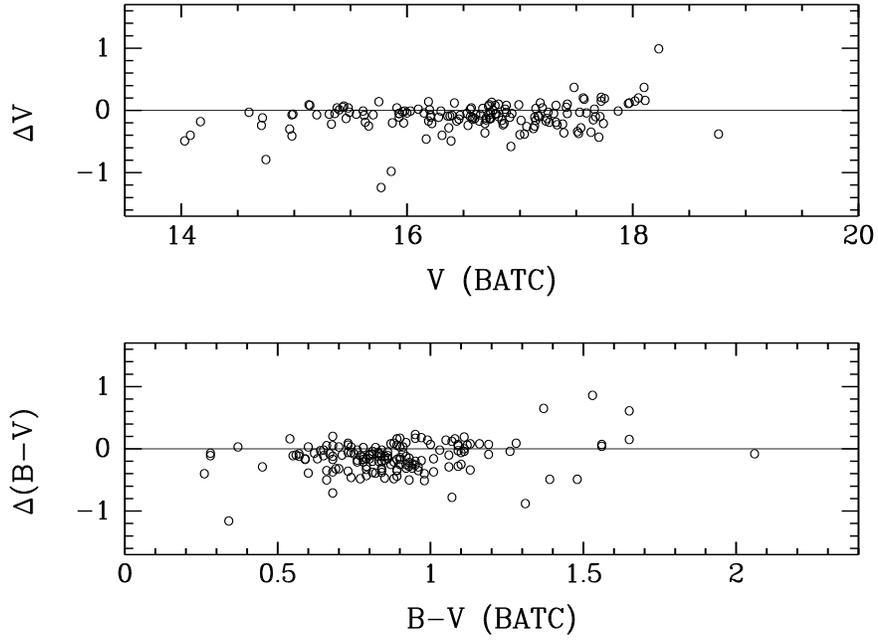,width=15.0cm,angle=-90}} 
\vspace{-1.0cm}
\caption{Comparison of GC photometry with Battistini et al. (1987). 
The vertical axis is the BATC photometry 
minus the measurement of Battistini et al. (1987).}
\end{figure}

\setcounter{figure}{2}
\begin{figure}[ht]
\centerline{\epsfig{file=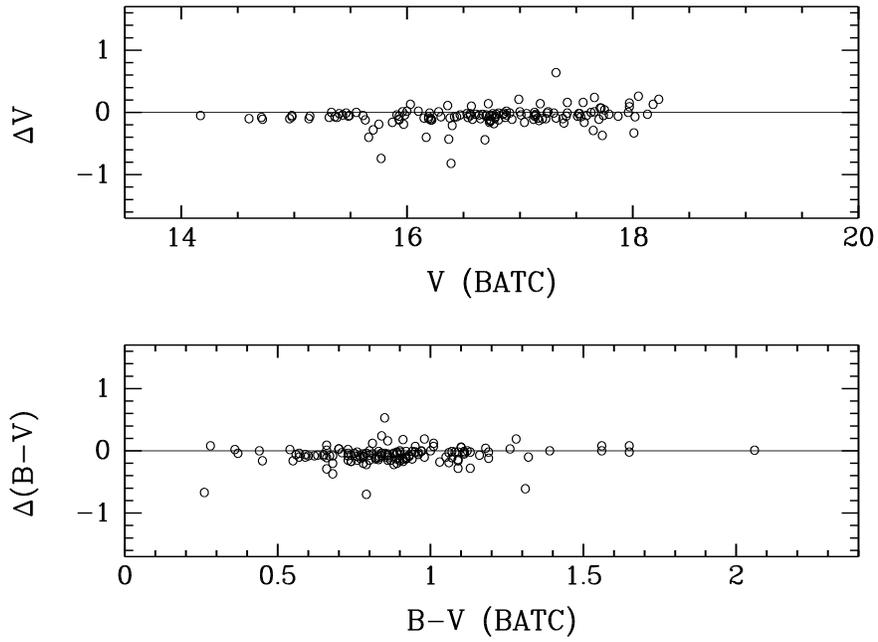,width=15.0cm,angle=-90}} 
\vspace{-1.0cm}
\caption{Comparison of GC photometry with Barmby et al. (2000). 
The vertical axis is the BATC photometry
minus the measurement of Barmby et al. (2000).}
\end{figure}                          


\setcounter{figure}{3}
\begin{figure}[ht]
\centerline{\epsfig{file=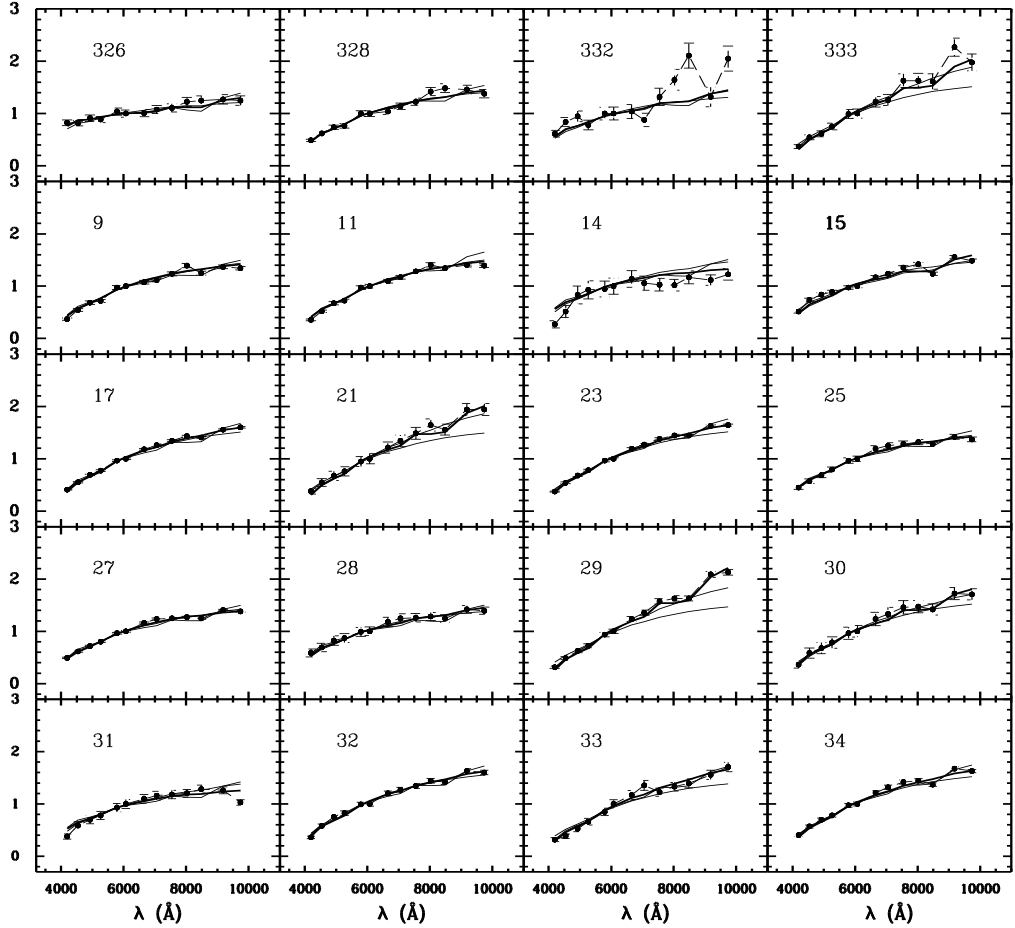,width=18.0cm,angle=270}}
\caption{Map of the fit of the integrated color of a SSP (Z = 0.0004, 0.004, 
and 0.02) with intrinsic integrated color for 20 GCs selected from the 172 GCs 
(the first 20 GCs in Table 2). Filled circle represents the intrinsic 
integrated color of a GC, the thick line represents the best fit of the
integrated color of a SSP of GSSP, 
and the thin lines represent the other two fits. Y-axis is the ratio of 
the flux in each filter to the flux in filter BATC08.}
\end{figure}

\setcounter{figure}{4}
\begin{figure}[ht]
\centerline{\epsfig{file=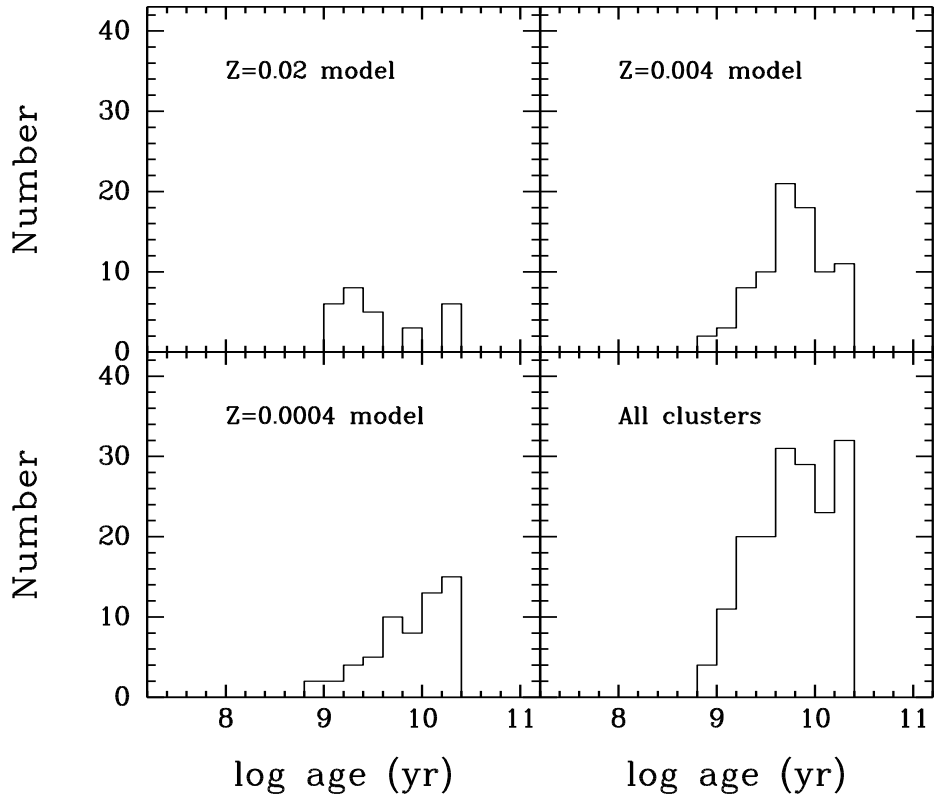,width=18.0cm,angle=270}}
\vspace{-1.0cm}
\caption{Histograms of M31 GC ages}
\end{figure} 

\setcounter{figure}{5}
\begin{figure}[ht]
\centerline{\epsfig{file=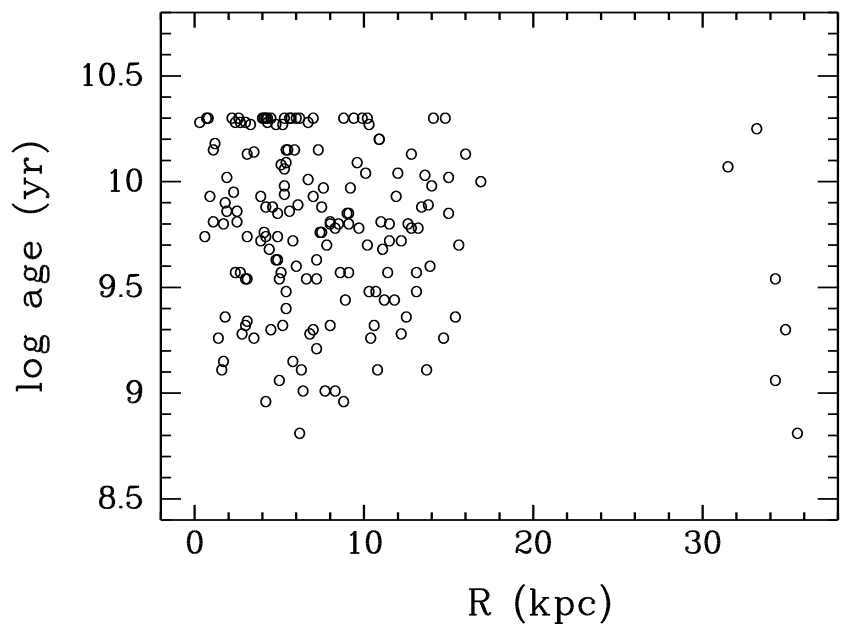,width=18.0cm,angle=270}}
\vspace{-2.0cm}
\caption{Age as a function of galactocentric distance for M31 GCs.
The absence of GCs with $20 < R < 30 kpc$ is a selection effect.}
\end{figure}

\setcounter{figure}{6}
\begin{figure}[ht]
\centerline{\epsfig{file=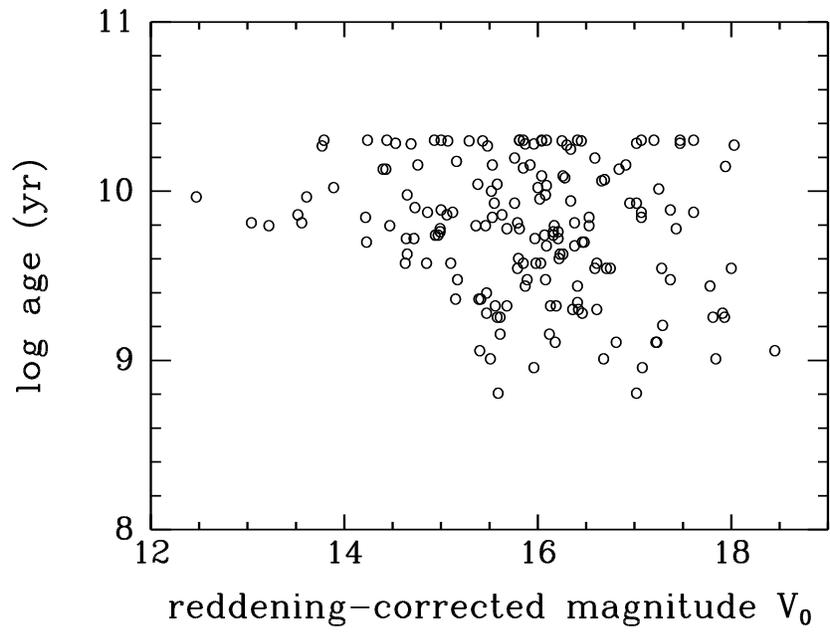,width=18.0cm,angle=270}}
\vspace{-2.0cm}
\caption{Age as a function of reddening-corrected magnitude $V_0$ for M31 GCs}
\end{figure}


\end{document}